\documentclass[prl,twocolumn,showpacs]{revtex4}
\usepackage{amsmath,amsfonts,amssymb,bm}
\usepackage{graphicx}
\usepackage{longtable}
\begin{document}
\bibliographystyle{apsrev} 
\title{ 
Patterson Function from 
Low-Energy Electron Diffraction Measured Intensities 
and Structural Discrimination 
}
\author{Celia Rogero, Jose-Angel Martin-Gago} 
\author{Pedro L. de Andres} 
\email{pedro@mailaps.org}
\affiliation{Instituto de Ciencia de Materiales (CSIC), 
Cantoblanco, 28049 Madrid, Spain}
\date{\today}
\begin{abstract} 
Surface Patterson Functions have been derived by direct inversion of 
experimental Low-Energy Electron Diffraction I-V spectra measured at 
multiple incident angles. The direct inversion is computationally 
simple and can be used to discriminate between different
structural models. (1$\times$1) YSi$_{2}$ 
epitaxial layers grown on Si(111) have been used to 
illustrate the analysis. 
We introduce a suitable R-factor for the Patterson 
Function to make the structural discrimination
as objective as possible. 
From six competing models needed to complete the 
geometrical search, four could easily be discarded, 
achieving a very significant and useful reduction in the parameter space
to be explored by standard dynamical LEED methods. 
The amount and quality of data needed for this analysis 
is discussed. 
\end{abstract}

\maketitle
Low-Energy Electron Diffraction (LEED) takes advantage of strong 
interactions between LEED electrons and atoms to achieve 
highly sensitive and accurate surface structure 
determinations\cite{vanhove:86}. 
The prize to pay for that sensitivity based on a strong interaction 
is that simple first-order 
perturbation techniques are not enterily appropriated, 
and multiple-scattering 
is unavoidable to analyze the experimental data. 
Because multiple-scattering implies 
self-consistency\cite{pendry:74}, 
direct inversion of data to obtain the 
structure is quite difficult, 
and the most popular route to surface structural 
determination relies on a trial-and-error procedure, that becomes 
too demanding when 
the parameter space to be explored is too big, or the structure 
cannot easily be guessed beforehand. 
Therefore, there is a genuine interest in developping and 
exploring direct inversion procedures, 
regardless of small practical difficulties or some limitations 
that might be found associated with them\cite{adams:77}. 
For LEED, an already succesful inversion technique based in ideas 
derived from 
holography exists\cite{saldin:90,heinz:01}. 
An useful alternative to holographic ideas, based in the Patterson 
Function (PF) concept,
has also been proposed in the literature by Tong et al.
\cite{tong:01}. These two proposals are based in different physical 
concepts, but have in common that they yield useful information 
on the surface structure by directly performing on the experimental data a 
Fourier transform-like with an appropriate kernel. 
The main desired effect of this mathematical operation is to
filter out high order scattering terms. Out of necessity, this
is only an approximate procedure, but previous experience with
holography, iterative methods based in maximum entropy\cite{maxent:02},
or even quasi-dynamical methods\cite{bickel:85} have shown that
multiple-scattering is not necessarily a roadblock.

In this paper we explore the PF idea, and by considering a practical example 
based in experimental LEED-IV data measured at different energies and 
incident angles we show how useful these ideas can become when they 
are used to complete a real surface structure determination. 
The strategy to defeat the limitations imposed by multiple-scattering 
is to collect data simultenously at different energies and incident angles. 
This results in overdetermination of the inverted data, where 
coincidences in features not directly related to the 
underlaying geometrical structure 
decrease as the data base increase\cite{deandres:92}. 
Therefore, the effort to get the 
structure shifts from the trial-and-error search to the measurement part,
which
is not a real burden taking into account 
modern LEED data acquisition techniques.

The surface Patterson function (PF) 
can be computed by an appropriate phase sum of the 
measured intensities\cite{tong:01}:
\begin{equation}\label{eq:PF} 
P(\vec r) = \mid \sum_{\vec k_{i}} \sum_{\vec g_{\parallel}} 
\int I(\vec k_{i}, \vec q) e^{i \vec q \vec r} d q_{\perp} 
\mid^{2} 
\end{equation}
\noindent 
where $\vec k_{i}$ is the wavevector corresponding to incoming 
electrons, $\vec q$ represents the momentum transfer for outgoing 
electrons detected in the LEED screen 
($\vec k_{f}= \vec k_{i} + \vec q$), 
and $\vec g_{\parallel}$ labels the different beams. 
Simple real-space scattering 
models analyzed by Tong et al.\cite{tong:01} help us to understand why the 
multi-energy and 
multi-angle approach can reduce spurious features appearing in 
the stationary phase condition implied in Eq.~(\ref{eq:PF}). 
The PF yields maxima at positions related to interatomic distances: 
we shall find that these can be used to discriminate reasonable from 
unreasonable structural candidates. 

\begin{figure} 
\begin{center} 
\includegraphics[width=1.\columnwidth]{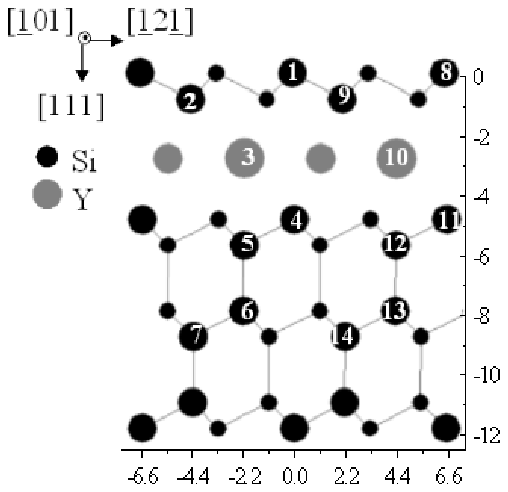} \\ 
\includegraphics[width=0.99\columnwidth]{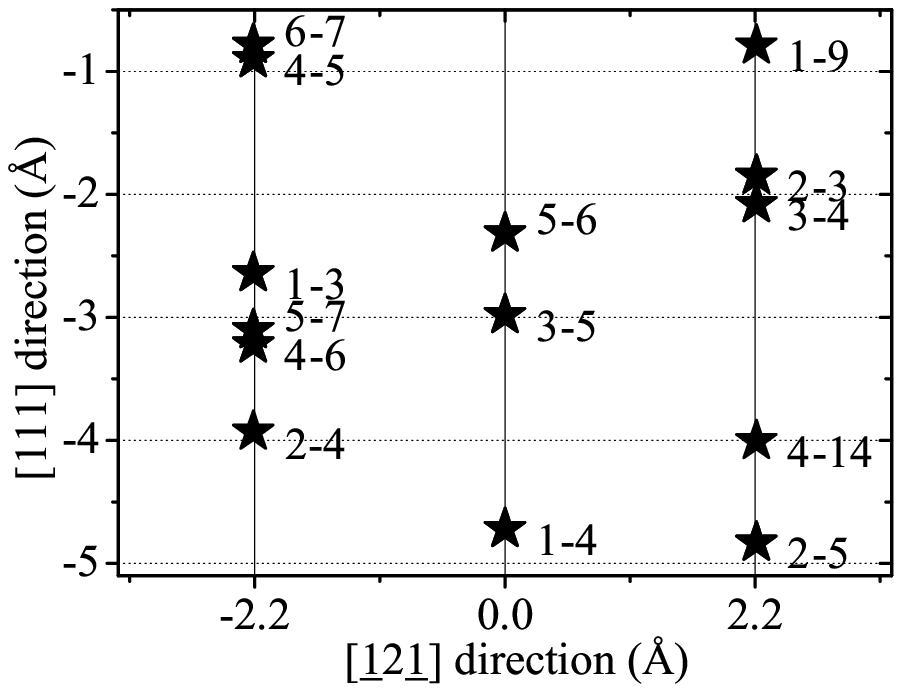} 
\end{center} 
\caption{\label{geom} 
(A) Upper pannel: 
Schematic positions and labelling of atoms in YSi$_{2}$/Si(111). 
The geometry corresponds to the best fit determined to experimental 
LEED data measured on the same system\cite{rogero:02} 
(B) Lower pannel gives the geometrical Patterson 
Function related to structure in upper pannel. 
} 
\end{figure}

To better understand how useful these ideas might be for real structural 
determination we take advantage of the experimental LEED 
I(V) curves we have recently measured at normal and non-normal 
incidence for 
(1$\times$1) YSi$_{2}$ layers grown on Si(111) substrates. 
These measurements have been used to 
perform a standard I(V) study of the structure, where the best correlation 
was found at a Pendry R-factor\cite{pendry:80} 
of $R_{P}=0.21$, which is accepted to be a low-enough 
value to credit the structural model. Details related to this model 
have been published elsewhere\cite{rogero:02}, but the structural model is 
presented in Fig.~(\ref{geom}A) for reference. Basically, the surface is made of 
Si(111) planes $180^{\circ}$ rotated around the Si(111) surface normal, 
with Y atoms located below the surface on T$_{4}$ sites (w.r.t 
the Si(111) substrate). The small relaxations 
between layers included in the model are 
not crucial for the analysis presented in this paper, because the spots 
on the PF already have an intrinsic width on the same order or 
bigger than the values associated to the pure geometrical 
surface relaxations.

Experimental LEED I(V) curves were recorded at normal 
and non-normal incidence. 
The latter was achieved by moving the surface normal in a plane 
defined by the surface normal and the 
$\lbrack \overline{1} 2 \overline{1} \rbrack$ surface 
direction. Measurements at $\theta= 0^{\circ}, 4^{\circ}, 9^{\circ}$ 
and $14^{\circ}$ where taken for the same azimuthal angle, $\phi$. 
19 beams (6 of which are not symmetry-related) where measured from 50 
to 450 eV, except for normal incidence where only 18 beams could be 
collected (the specular beam was blocked by the electron gun). 
Therefore, the total energy range for independent data is more than 
$8000$ eV. The experiment was performed at room temperature: in the standard 
LEED analysis it was found that fine tuning of isotropic vibrations 
associated to different layers could make a strong impact on the R$_{P}$ 
value, decreasing its value from around $0.4$ for r.m.s. displacements taken 
from bulk values, to around $0.2$ for the final optimized surface
 values\cite{rogero:02}. 
At the moment these effects are not taken into account in 
Eq.~(\ref{eq:PF}), but we anticipate their main effect is to broaden the PF 
increasing the values of the correlation factor we have found below. 

\begin{figure} 
\begin{center} 
\includegraphics[width=0.99\columnwidth]{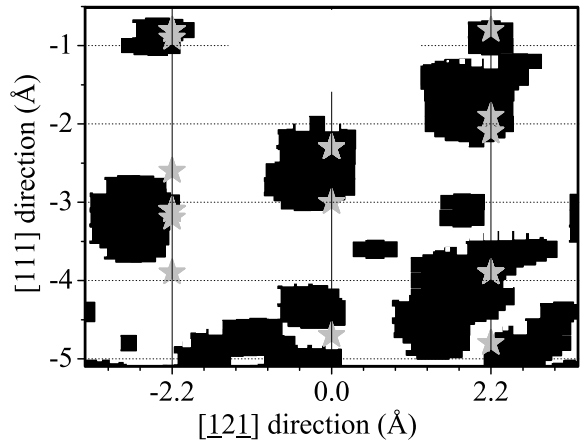} 
\end{center} 
\caption{\label{pattpatt} 
PF from 
Eq.~(\ref{eq:PF}), and the experimental I(V) 
curves measured for 19 beams at four different 
incident angles. 
} 
\end{figure}

The PF gives the interatomic distances between all atomic pairs in the 
structure. To simplify the discussion and make 
a better representation of 
our results, we have chosen the $\lbrack \overline{1} 0 1 \rbrack$ 
cross-sectional plane, but the same conclusions would be obtained using 
a different plane. Fig.~(\ref{geom}B) shows the interatomic distances 
for one unit cell 
calculated from the structure in Fig.~(\ref{geom}A): stars correspond 
to positions where a perfect stationary phase condition would
produce spots in the PF from I-V spectra. The 
corresponding atomic pairs are indicated next to the stars, using the 
same labelling as in Fig.~(\ref{geom}A). 
This is a mere geometrical PF, and it is the best result that could ever
 be recovered for a given structure. 
Therefore, we compute
correlations between 
experimental PF obtained from 
Eq.~(\ref{eq:PF}) using all measured diffracted intensities
and geometrical PF corresponding to different trial structures.
As shown in Fig.~(\ref{pattpatt}), the experimental 
PF is made of spots with a certain width derived from factors like 
the limited domain of integration, atomic vibrations, defects, etc.

A visual comparison between the stars in Fig.~(\ref{geom}B) and 
the spots in Fig.~(\ref{pattpatt}) induces 
us to introduce a correlation factor 
between PF's to facilitate their comparison. 
The new correlation factor might be defined in different ways, 
but to demonstrate the concept  
we shall here introduce the simplest one, to keep the discussion as 
independent as possible from particular details of this definition:
\begin{equation}\label{RPF} 
R_{PF} = \frac{S_{spots}}{S_{PF}} \frac{\sum D }{N_{geo}} 
\end{equation}
\noindent 
where $D$ is the distance between stars 
corresponding to a given structure
and the center of the closest experimental spot around it, 
$N_{geo}$ is the number of interatomic distances considered, $S_{spots}$ is the area covered 
by the spots, and $S_{PF}$ is the total area considered for the analysis. 
The factor $\frac{S_{spots}}{S_{PF}}$ gives the probability to find a spot in 
the considered area, whilst the factor $\frac{\sum D}{N_{geo}}$ measures 
how far are the experimental spots 
from the corresponding stars. In order to preserve 
the meaning 
of this particular $R_{PF}$, 
the factor $\frac{S_{spots}}{S_{PF}}$ should not be 
too close to $1$ (e.g., spots covering the whole surface) or to $0$ 
(e.g., too few spots). By constructing model geometries at random, 
we estimate
typical values for $R_{PF}$ 
to be around 0.25-0.20
when the structures are not correlated.

The structural analysis for YSi$_{2}$ epitaxially grown on Si(111) 
requires considering six different models for different stackings 
of the atomic layers. 
In our previous analysis of this structure\cite{rogero:02}, 
a detailed R$_{P}$ analysis covering different values for all 
the considered parameters has been performed. This extensive search 
reflects the sensitivity and highly non-linear dependence 
of intensities with structural parameters in multiple-scattering 
calculations\cite{deandres:00}. 
This is not the case for PF, where a stationary phase condition 
based in two first-order scattering events is sought. Therefore,
the advantage of computing correlations between experimental PF 
and different models is double: 
(i) there is no need to run costly 
multiple-scattering calculations for the whole parameter space, and 
(ii) because the analysis is based in first-order scattering
events,
the R$_{PF}$  is expected to behave smoother as a function of
the involved parameters, and less local minima are expected.
Although, on the other hand, 
a smoother variation may result in less sensitivity
to different structures.
Correlations between the experimental 
PF and the different models considered in the previous structural analysis are 
given in table ~\ref{tab:6mod}. 
To complement R$_{PF}$, we give next to it 
a number counting the percentage of coincidences between the
stars and the PF spots. This has been defined as
the number of stars which are inside the region in black in
the figures.
This number can be helpful to discriminate between models,
as can been seen in Table~\ref{tab:6mod},
but it has not been included in the
R$_{PF}$ definition because it depends on the width of the 
spots (a function of several factors, e.g. the pure thermal
vibration of atoms is expected to contribute an intrinsic width
in the order of several tenths of Angstroms). 
 
\begin{table*} 
\caption{\label{tab:6mod} 
$R_{PF}$ and coincidences for different stacking models. Results are 
shown for the experimental data and two theoretical simulations 
(19 and 91 beams). Coincidences give the 
percentage of geometrical stars inside  
experimental or theoretical PF-spots. 
} 
\begin{ruledtabular} 
\begin{tabular}{|c|c|cc|cc|cc|} 
Code & Structural Model & \multicolumn{2}{c|}{Experiment}& \multicolumn{2}{c|}{Theory (19 beams)}&\multicolumn{2}{c|}{Theory (91 beams)}\\ 
 & & R$_{PF}$ & (\% Coinc.)& R$_{PF}$ & (\% Coinc.)& R$_{PF}$ & (\% Coinc.)\\
\hline 
A & T$_{4}$ Rotated 0$^{\circ}$ & 0.159 & 40 & 0.114 & 60 & 0.113 & 67 \\ 
B & T$_{4}$ Rotated 180$^{\circ}$ & 0.140 & 50 & 0.069 & 88 & 0.026 & 88 \\ 
C & H$_{3}$ Rotated 0$^{\circ}$ & 0.175 & 40 & 0.122 & 60 & 0.146 & 53 \\ 
D & H$_{3}$ Rotated 0$^{\circ}$ & 0.150 & 47 & 0.123 & 59 & 0.060 & 65 \\ 
E & Atop Rotated 0$^{\circ}$ & 0.161 & 50 & 0.088 & 73 & 0.107 & 86 \\ 
F & Atop Rotated 180$^{\circ}$ & 0.136 & 67 & 0.083 & 73 & 0.030 & 80 \\ 
G & random model & 0.250 & 13 & 0.250 & 13 & 0.180 & 6 \\
\end{tabular} 
\end{ruledtabular} 
\end{table*}

From entry named {\it Experiment} in 
table~\ref{tab:6mod} we can inmediately reduce the available six models 
to two (i.e., a 66\% saving has already been achieved by 
simply Fourier-transforming 
the experimental data). Differences in $R_{PF}$ 
between model B and F, however, are not big enough to discriminate 
between them.
In fact, 
the lowest $R_{PF}$ corresponds to model F, 
whilst the best fit based in $R_{P}$ corresponds to the second lowest 
$R_{PF}$-value, model B. 
Notice that in both cases, the percentage of 
coincidences are among the highest.

To understand the origin of this different ordering in models based 
in PF analysis, 
we test the consistency in the definition of $R_{PF}$ 
by replacing the experimental I(V) curves by 
theoretical ones corresponding to the structure B. Results shown in 
table~\ref{tab:6mod} under the column named {\it Theory (19 beams)} 
show that the definition for the R-factor is internally consistent and now 
the best model is clearly the correct one. Therefore, the question arises 
of what is the effect on the PF of exchanging sets of I(V) curves (i.e., 
experiment and theory) that differ themselves by $R_{P}=0.21$. 
We notice that multiple-scattering cannot be the source for the discrepancy
as both, the experimental and the theoretical intensities include all
the multiple-scattering relevant for the problem: more likely this is due to
residual factors (like anisotropic vibrations of atoms
near the surface, defects, etc) or the size of the integration domain itself.
\begin{figure} 
\begin{center} 
\includegraphics[width=0.99\columnwidth]{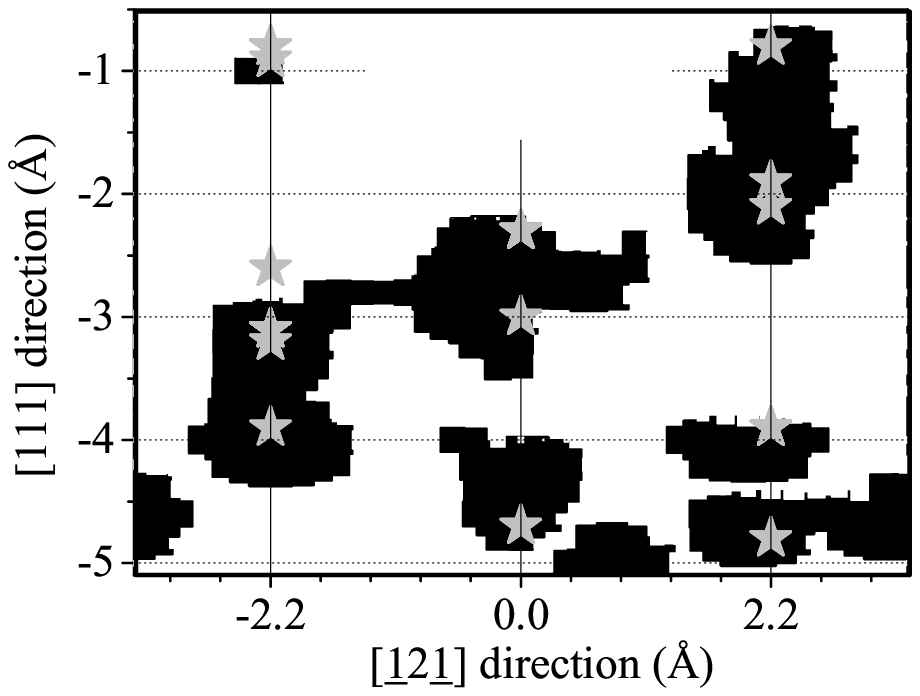}\\ 
\includegraphics[width=0.99\columnwidth]{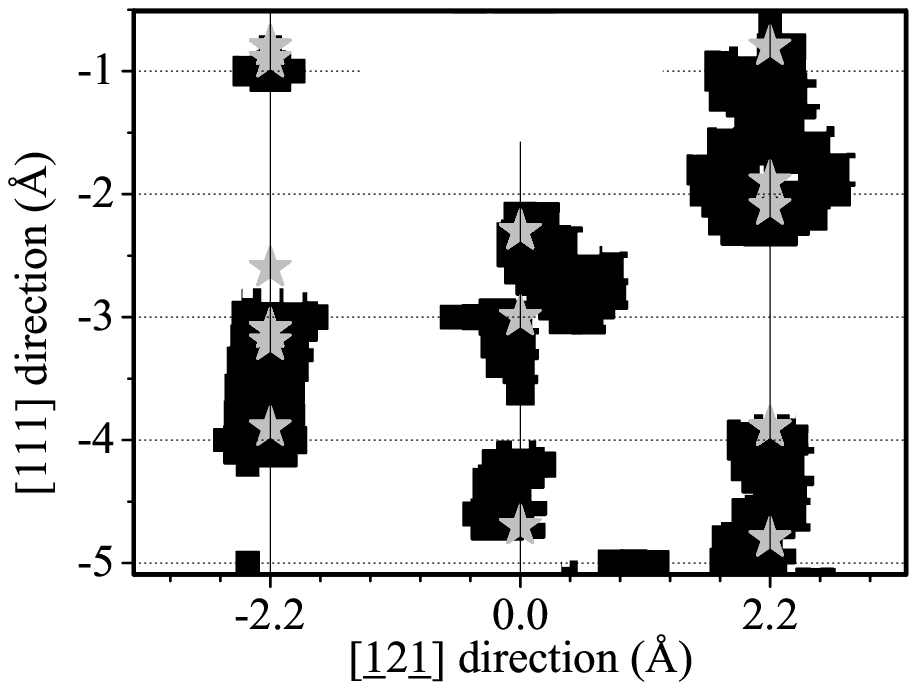} 
\end{center} 
\caption{\label{pattpattTH} 
(A) Upper pannel shows the 
PF obtained from Eq.~(\ref{eq:PF}), and the theoretical I(V) 
curves for 19 beams at four different 
incident angles computed for geometry in Fig.~(\ref{geom}). 
(B) Lower pannel shows the 
PF obtained from Eq.~(\ref{eq:PF}), and the theoretical I(V) 
curves measured for 91 beams at four different 
incident angles. 
} 
\end{figure}
Therefore, we turn to a
visual comparison of Fig.~(\ref{pattpatt}) and 
Fig.~({\ref{pattpattTH}A). Both PF's have a
similar probability to find a spot: 
$\frac{S_{spots}}{S_{PF}}=0.40$ for experiment and 
$\frac{S_{spots}}{S_{PF}}=0.39$ for theory, 
but widths of individual spots are too wide 
to obtain an accurate correlation with stars related to models B and F.
Due to spurious features, both models become undistinguishable:
this is not too surprising as the two models in question are very similar.

Two simple ways to narrow these widths come to mind. 
From an experimental point of view, measurements made at low temperatures 
or recorded on samples as perfect as possible are needed. 
On the other hand, the stationary phase condition
implied in Eq.~\ref{eq:PF} requires covering 
as much volume as possible in the 
Fourier-transform domain 
(i.e., measuring as many beams at different angles for as many
energies as possible). 

Our experimental setup puts some limits on the number of beams we can 
measure, and restrict us to room temperature. 
We comment in passing that analyzing a $1 \times 1$ put us in the worst 
possible case for PF analysis. For the same energy range, any other 
superstructure will provide extra  beams to be measured.
Furthermore, the $1 \times 1$ usually would contribute less stars
than a system with a surface superstructure,
where different pair distances 
are involved. Finally, beams in the $1 \times 1$ mix
surface and bulk information, a disadvantage
that the fractional beams corresponding to a genuine
surface superstructure do not present.

Therefore, we explore the second requirement 
by substituting our experimental I(V) curves by theoretical 
I(V) curves computed for model B for all the possible exiting 
beams (i.e., 91 beams in total). 
Indeed, comparison between Figs.~(\ref{pattpattTH}~A) 
and ~(\ref{pattpattTH}~B) demostrates 
that the widths of the PF spots narrow when the data base is increased: 
the probability to find a spot 
has been reduced to $\frac{S_{spots}}{S_{PF}}=0.25$, 
nearly a factor of two w.r.t. the theory or the experiment based in 19 beams.
Clearly,
if we can measure more beams we could obtain more accurate
correlation between stars and PF-spots.

In conclusion, we have found that the PF concept allows us to 
significatively cut 
the parameter space to be explored in a real structure determination based 
in LEED data measured at different angles and energies. 
The extra burden of having to measure more beams and 
the requirement for liquid Nitrogen cooling temperature 
should not be excesive for modern LEED equipment.
The advantage is significative. 
The analysis based in the PF 
is very fast compared with the standard procedure, 
as it mainly implies Fourier-transforming 
the measured data. 
Inside the restricted parameter space determined 
by the PF analysis, it seems appropriate to 
run multiple-scattering calculations. 
This is certainly more costly than the simple integral 
involved in PF but it might be worthwhile:
its intrinsic non-linear dependence 
with the structure, and the presence of higher order
correlations, will help to obtain accurate 
and trusty values for the final parameters of the model.

Work supported by the Spanish 
Ministry of Science and Technology (PB98-524).

\small

\bibliography{patterson}

\end{document}